# A Computer Vision Aided Beamforming Scheme with EM Exposure Control in Outdoor LOS Scenarios

Tianqi Xiang*, Huiwen Li*, Boren Guo*, Xin Zhang*
*Wireless Theories and Technologies Lab
Beijing University of Posts and Telecommunications
Beijing 100876, China
Emails: xiangtianqi@sina.com; lhw5799@bupt.edu.cn; plume@bupt.edu.cn; zhangxin@bupt.edu.cn

*Abstract*—Without any radiation control measures, a large-scale mmWave antenna array at close range may lead to a large amount of electromagnetic exposure of human. In this paper, with the aid of pose detection in computer vision, a beamforming scheme using a novel exposure avoidance method is proposed in outdoor line of sight scenarios. Instead of reducing transmitted power, the proposed method can protect the vulnerable parts of human body from electromagnetic exposure during transmission by deviating the transmission beams from vulnerable parts. Besides, a finer beam management granularity is adopted to better balance the trade-off between exposure reduction and communication quality loss, because finer beams can provide more adjustability for finding the beam that reduces exposure without excessively reducing the link quality. The proposed exposure avoidance method is validated in simulations, and the results show that the finer beam management granularity can guarantee communication quality while reducing the electromagnetic exposure.

## I. INTRODUCTION

Radio frequency electromagnetic (EM) exposure has always been a concern in the development of mobile communication systems, and mmWave in 5G is no exception. According to the standard of exposure limits specified by ICNIRP[1], IEEE[2] and FCC[3], most of the current researches focus on the exposure to mobile devices[4][5][6], and some of them focus on the exposure to macro base station[6] or micro base station[4].

Meanwhile, in the assessment modeling of EM exposure, the beamforming technology characterized by directional wireless communication should also be considered[6]. Beamforming technology is used to improve signal coverage quality, especially suitable for mmWave band featured by large bandwidth and high propagation loss. And with the increase of antenna array scale, the expected beamforming gain rises. However, the energy of base station antenna array is also concentrated, more likely leading to severe EM exposure problems, especially if people are quite near the BS. For example, researchers in [7] have built an indoor large-scale AAS mmWave communication prototype, in which a considerably high equivalent isotropically radiated power (EIRP) is achieved in a very short distance, which might need better exposure control. But as far as the author's knowledge is concerned, there has been no relevant literature, including [7], focusing on the radiation control measures of large-scale antenna array in such scenario.

However, beamforming or precoding technology could have inherent advantages in reducing exposure. Researchers in [8] use precoding to reduce exposure, where they achieved a spatial distribution change of signal energy by applying precoding, so that the exposure of human body is artificially brought down. We expect that this can also be achieved with beamforming. But how to 'passively', or in a manner with less radio channel overhead, acquire the spatial channel information or the gesture/position of human body still remains as a challenging issue.

In our previous work[9][10], we used computer vision (CV) as the pilot to detect the position of pedestrians in the outdoor LOS scenarios and select beams which are pre-generated, while preserving users' privacy as much as possible. This sheds some light on how to address the EM exposure issue in a better way. As wireless communication system itself can also be used for human body perception[12], but compared with CV, wireless detection or perception of human body will consume more wireless resources to illuminate objects (probably with higher EM exposure) and will be difficult to be compatible with the existing mobile communication systems.

In addition, similar to [10], we consider the outdoor close-range mmWave coverage scenario, assuming that the antenna array of base station is large and the UE antenna is a single omni-directional antenna. On the one hand, pedestrians at close range will obtain not only high coverage quality but also probably high exposure risk from the antenna array of base station. On the other hand, this assumption indicates low requirements for UE so that UE does not consume too much computing resources and transmitted power, which is beneficial for UE to save energy and reduce radiation from UE to human body respectively. Therefore, this paper addresses reducing EM exposure to base station without excessive degradation of link quality obtained by large-scale antenna array.

In this paper, we propose a beamforming scheme considering EM exposure control in LOS scenarios with the aid of CV. An exposure avoidance method based on beamforming is proposed, which makes the transmission beam deviate from the vulnerable parts (such as the head) of pedestrian in angular space and aim at UE as much as possible under the condition that the locations of UE and the head of pedestrian can be detected by CV. Because mmWave beams with large-scale antenna array have good directivity, beams are able to distinguish pedestrian's head and UE in angular space, thus resulting in exposure reduction. Meanwhile, in order to meet the positioning requirements of the proposed method, we apply pose estimation[11] in CV to extract the key joints of pedestrians, and further extract the positions of potential terminals and important parts of human body.

Then we note that there is a trade-off between reducing exposure and ensuring communication quality when exposure avoidance is triggered, because the beam suggested by exposure avoidance method may not be the optimal beam for transmission. And in the normal mmWave beam management of NR[13], where the beam spacing is defined by half power angle, the beam management granularity is not fine-grained enough that the exposure avoidance may lead to certain loss of communication quality at the same time. Therefore, we apply finer beam management granularity to provide more adjustability for the trade-off between exposure reduction and communication quality assurance.

The organization of the paper below is as follows. Section II introduces the optimization problem and the exposure avoidance method. Section III discusses the beam management granularity. Section IV gives the performance metrics and section V includes the simulation results and analysis. Some future work is proposed and discussed in Section VI and section VII makes a conclusion of this paper.

## II. EXPOSURE AVOIDANCE

In wireless communication systems considering the limitation of electromagnetic radiation to human, we assume an optimization problem, which is to improve the link quality as much as possible while ensuring that the exposure does not exceed the specified value. However, these two aspects are generally contradictory, because improving the link quality by increasing transmitted power may increase the EM exposure in the environment as well. And when considering the limitation of exposure in downlink for example, it is generally required that the transmitted power of base station is not higher than a specific value, which, however, is not conducive to the optimization of link quality.

Beamforming technology brings spatial distinction of electromagnetic energy. Moreover, when a large-scale antenna array is equipped, the change of spatial electromagnetic energy can be sufficiently differentiated so that the spatial dimension could be one of the dimensions for radiation control. When the transmitted power is not strictly limited, some beams may radiate the vulnerable parts of human body exceeding the exposure limit. We disable these beams for data transmission. Although some beams are disabled, the optimal beam (when it is not disabled) or sub optimal beam (when the optimal one is disabled) can still provide high link quality.

In order to obtain beams that may lead to excessive exposure, feedback or prediction mechanism is needed. For example, pedestrians can wear sensors or use mobile phones for radiation measurement and feedback, but this will take up more wireless overhead and higher cost. Or the radiation value of each grid in the environment can be measured in advance and be referred for beam suggestion. But this demands highly accurate positioning, and with smaller grids and more beams, the data scale will be larger. In LOS scenarios, CV can be used to predict the relative strength of the signal because the signal power of the main path takes up main component of total power. And when the communication terminal or human body is close to the peak of the main lobe of a transmission beam, there will be higher beamforming gain or exposure, and vice versa. Therefore, in the LOS scenarios, CV could be applied to control exposure by predicting exposure and signal strength, as presented in Figure 1.

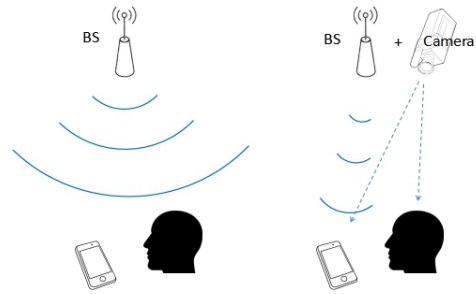

(a) uncontrolled radiation  (b) radiation control with CV

Figure 1. Illustration of uncontrolled radiation and radiation control with CV

Then we introduce the exposure avoidance method which aims to point the transmission beam at the communication terminal as closely as possible when the important parts of human body are less radiated. We call it exposure avoidance because beams avoid pointing directly at vulnerable parts of the human body, thus reducing exposure, no matter the human is using a mobile phone or not. When exposure avoidance becomes necessary while data transmission, visual detection system will acquire positions of important parts and communication terminals relative to antenna array, both of which can be extracted with pose detection in CV.

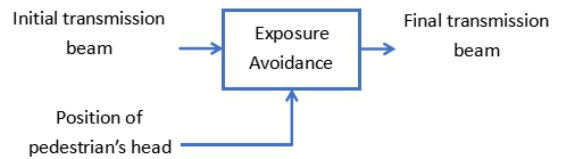

Figure 2. Mechanism of exposure avoidance. The initial transmission beam can be determined by CV[10] or beam management process[13].

Figure 2 shows the mechanism of exposure avoidance. Considering exposure avoidance of human head, this method recommends beams as the following formula:

$$\text{Beam} = \underset{}{argmin} \ d(\text{Beam}_{UE}, \text{Beam})$$
$$s.t. \ d(\text{Beam}_{Head}, \text{Beam}) \geq d_0 \quad (1)$$

wherein $\text{Beam}_{UE}$ is the initial transmission beam ID and $\text{Beam}_{Head}$ is the beam closest to human head which could result in the highest exposure. The variable Beam, $\text{Beam}_{UE}$ or $\text{Beam}_{Head}$ are selected from the codebook-based generated beam set[13][14]. Also as shown in Figure 2, $\text{Beam}_{UE}$ can be the transmission beam determined by the traditional beam management process[13]. In this case, exposure avoidance could be an additional function of 5G base stations equipped with cameras, similar to open-loop power control of mobile devices. We define $d$ as the distance operator of two beams, calculated as:

$$d(\text{Beam}_1, \text{Beam}_2) = \sqrt{(m_1 - m_2)^2 + (n_1 - n_2)^2} \quad (2)$$

wherein $m_1$ and $m_2$ are the index of $\text{Beam}_1$ and $\text{Beam}_2$ respectively in azimuth direction and $n_1$ and $n_2$ are the index in downtilt direction. $d_0$ represents the avoidance distance, which is arbitrarily given according to potential exposure risk. In order to evaluate the exposure quantitatively, the prediction method in [3] is used in this paper. Generally, without emission control measures, regions with high exposure risk are the regions close to antenna array. So the $d_0$ configuration is mainly based on the distance, that is, we set higher value of $d_0$ for closer regions. And when the distance is big enough thus with low exposure, $d_0$ can be zero, which means exposure avoidance will not be triggered.

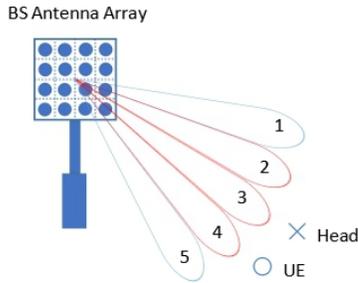

Figure 3. Illustration of how beams are forbidden and reselected.

Briefly, the exposure avoidance method in this paper can be interpreted as: determining the forbidden or disabled beam set according to the position of the human head and the risk level of exposure, and selecting the best communication beam with the knowledge of the disabled beam set. Figure 3 shows the process of disabling and reselecting. For example, when $d_0 = 2$, forbidden beam set includes beam 2,3 and 4. And the final transmission beam is beam 5 instead of the initial beam 4. It should be noted that the pedestrian head can be any head joints detected by CV, and does not necessarily belong to the pedestrian whose devices are being served by the BS. The exposure risk level is determined by the distance between the pedestrian and the antenna array. And the closer the distance is, the larger the disabled beam set is. And in the following, we call the exposure avoidance "triggered" when the initial beam is in the disabled beam set and thus the transmission beam needs to be reselected.

### III. BEAM MANAGEMENT GRANULARITY

Furthermore, this paper regards reducing exposure and ensuring communication quality as a trade-off when exposure avoidance is triggered. This trade-off is most obvious when the angle between the terminal and antenna array is similar to that between the human head and antenna array. In this case, the initial transmission beam is likely to be disabled due to exposure avoidance. And only a deviated beam can be selected as the transmission beam, which will lead to a decline of communication quality. Even if the UE and pedestrian's head are quite separated in angular space, it is possible that the optimal transmission beam is also disabled due to a high exposure risk. This can be called "exposure limited."

In NR, the beamwidth is calculated at 3dB of the main lobe of a beam, which is also considered as the angular spacing between beams in azimuth and downtilt direction[13]. When exposure avoidance is triggered, though the exposure can be brought down, the loss of beamforming gain might be huge. So this trade-off seems irreconcilable under this configuration: either reducing exposure, or guaranteeing the quality of communication.

In order to deal with this trade-off better, this paper adopts a finer granularity of beam management for exposure avoidance: beams are denser, or in other words, more fine-grained, between which the spacing is no longer the half power angle, but a smaller value. As presented in Figure 4, we expect that this will provide more adjustability for optimization.

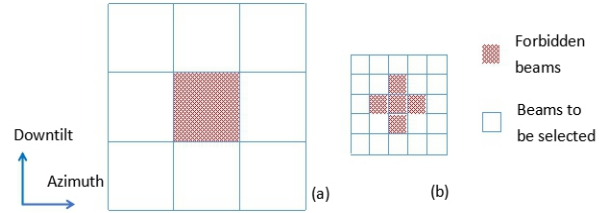

Figure 4. (a) less finer beam management granularity and $d_0 = 1$; (b) finer beam management granularity and $d_0 = 2$. A finer granularity promises more options for adjustment.

But when exposure avoidance is not triggered, the trade-off can be ignored. This is because the beam with good directivity can distinguish separated UE and head, or the exposure is originally low at far range. So it's not "exposure limited". Meanwhile, finer beam management granularity can attribute to more accurate beams and improve communication quality.

### IV. PERFORMANCE METRICS

In this paper, our simulation is mainly to verify the exposure reduction effect of exposure avoidance method. But in view of the trade-off between exposure reduction and communication quality, we will evaluate both of them at the same time.

The evaluation of exposure follows the prediction method in [3], only considering the exposure of pedestrian's head. In order to observe the effect of exposure avoidance method, this paper will simulate the situations with and without exposure avoidance method. And in order to investigate whether denser beams can play a better role in trade-off, our simulation includes two beam densities: 3dB and 0.5dB beam spacing.

Only single cell coverage is considered so far in this paper, so the broadband SNR is used to evaluate the communication quality, which is calculated as below:

$$\text{SNR} = \frac{P_T \cdot G \cdot PL}{P_{noise}} \quad (3)$$

where $PL$ includes pathloss and shadow fading based on [15]. $P_T$ is the transmitted power and $P_{noise}$ is the noise power. $G$ is the antenna array gain including element pattern and beamforming gain. For $N_V \times N_H$ elements, it is expressed as[14]:

$$G(\text{dB}) = G_E + 10\,log_{10}\left(\left|\sum_{m=1}^{N_H}\sum_{n=1}^{N_V} w_{n,m} \cdot v_{n,m}\right|^2\right) \quad (4)$$

wherein $G_E$ is the element pattern. $w_{n,m}$ is the weighting and $v_{n,m}$ is the position vector. Due to LOS scenarios, so far we only take the main signal path into account.

As for CV performance, the CV assisted beamforming scheme of [10] is used. However, this paper requires pose estimation rather than detection tasks which can distinguish joints of human body. In order to obtain the angular error of positioning using pose estimation, we convert the deviated distance normalized by head size[11] into angular deviation. Next, we apply the same method as [10] to introduce this error into the beam selection process.

In addition, this paper makes two assumptions about the pose of pedestrians using mobile devices, as presented in Figure 5. The first (pose A in the next section) is that the mobile device is close to the pedestrian's head, such as the pedestrian is answering the phone or wearing VR helmet or smart glasses. In this case, as long as pedestrians are in the exposure risk area, it is very likely to trigger exposure avoidance. The second (pose B in the next section) is that the mobile device is relatively far from the pedestrian's head, such as using a mobile phone in front of the body taking selfie or online chatting or wearing a smart watch. And in this situation, only when the head is on or close to the main path of signal propagation, can exposure avoidance be triggered.

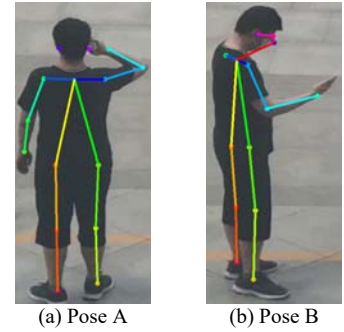

(a) Pose A    (b) Pose B

Figure 5.  Examples of two poses

## V. SIMULATION RESULTS

We first evaluate the performance of angular positioning error with CV. In [11], the researchers apply part affinity fields to multi-person pose estimation and obtain the average precision (AP) of key joints normalized by head size based on the MPII data set. We turn this into angular error of this paper's scenario, as shown in Figure 6.

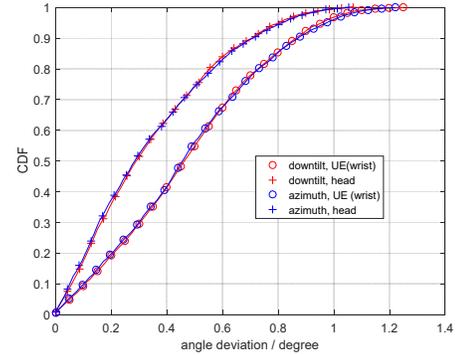

Figure 6.  Angular deviation obtained with CV

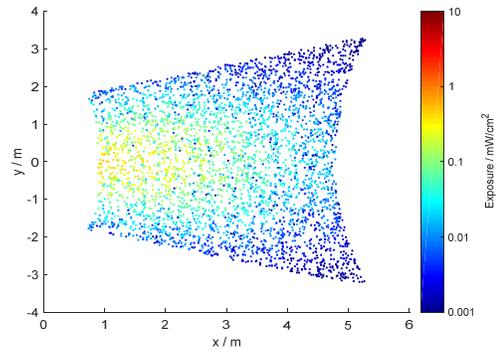

Figure 7.  Exposure risk map of pose A

We then assess the exposure risk of the scenario considered in this paper. Figure 7 shows the exposure map of pedestrian's head for pose A when no emission control measure is adopted. We apply similar assumptions as in [10], but the scale of antenna array is 32x32, and so the beams are denser and more accurate. The BS is at the origin and its height is 5 m. And the evaluation area in this paper is about

20 m$^2$. The transmitted power is 20 dBm, and due to the high array gain and the close range between pedestrian's head and antenna array (from about 3.5 m to 6.7 m), the power density of exposure is high, which indicates a great exposure risk. Since the UE and head is relatively close under pose A, the beam is directed at the head to some extent. So we also consider Figure 7 as the exposure risk of this paper's scenario and set $d_0$ based on it.

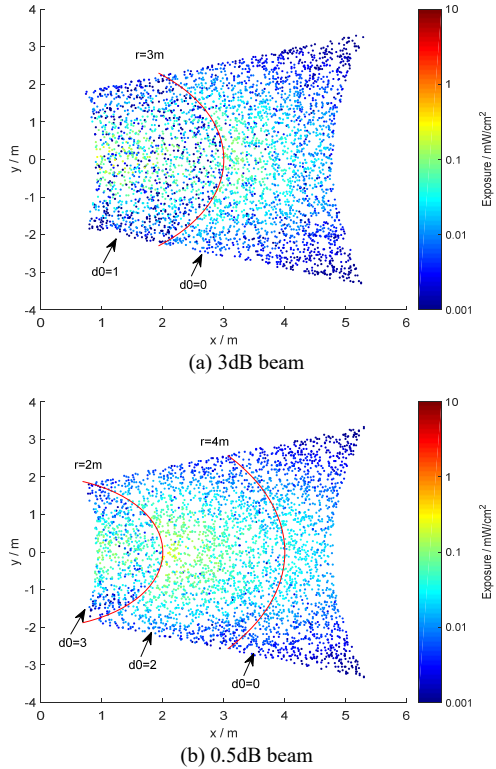

Figure 8. Exposure map of pose A applying exposure avoidance method

Figure 8 shows the $d_0$ strategy and exposure reduction effect for pose A of the two configurations. For 0.5dB beam, we use a stepped $d_0$ configuration and apply a larger value of $d_0$ at a closer region. However, for 3dB beam, we consider that too much deviation of a beam will lead to a great decline in communication quality, so there is only one step. And it can be seen from Figure 8 that when $d_0$ increases, the exposure will decrease, which indicates the effect of exposure avoidance method. Meanwhile, the exposure reduction effect of 3dB beam is more obvious than that of 0.5dB beam at the step crossing position, because the beam has a larger deviating angle.

Figure 9 shows that the exposure reduction effects of both 3dB and 0.5dB beam are similar, which basically keep the exposure within 0.3mW / cm$^2$ in this scenario.

Figure 10 shows the SNR CDF when exposure avoidance is applied. The loss of communication quality of 3dB beam is very serious under pose A, which makes exposure avoidance not worth the loss. However, the link quality of 0.5dB beam is slightly lower under pose A, and even performs better under pose B, because beams are more accurate and exposure avoidance is less triggered. And we note that under pose B, the contradiction between exposure avoidance and communication quality is relatively inapparent, in other words, less "exposure limited", because the pedestrian's head is less likely to be in the path of the optimal transmission beam. So for pose B, the well directional beams are able to distinguish them, as long as the beam management granularity is fine enough. So the exposure control of human body is realized while guaranteeing high gain of antenna array. We attribute this to the combination of exposure avoidance method and fine-grained beams.

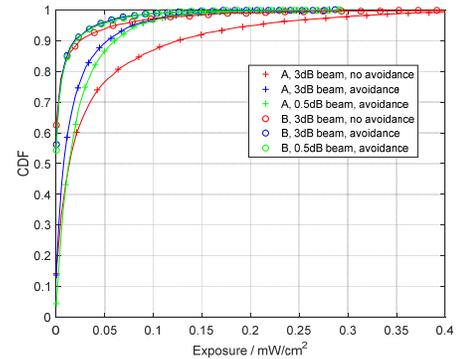

Figure 9. Exposure CDF

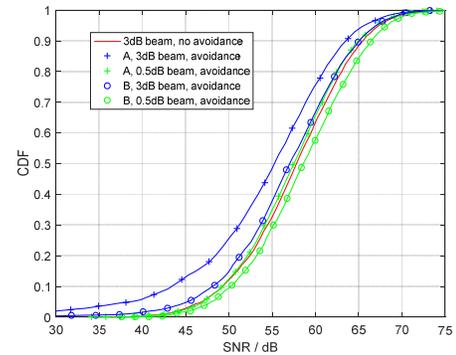

Figure 10. SNR CDF

The simulation results show that the spatial dimension could be an effective dimension for controlling exposure. While reducing the exposure to the specified value, communication quality can also be ensured, which won't be achieved by reducing transmitted power. Moreover, when the beam granularity gets finer, the proposed method performs better.

VI. FUTURE WORK

While some basic work has been conducted in this paper, there can be more research topics to investigate by applying the underlying idea of our work, which are briefly discussed as below.

Fast and accurate identification of the transmission state as LOS and NLOS is of great importance for 3D MIMO system[16]. And CV is able to obtain whether the object is blocked and the transition of two channel states (such as pedestrian in and out of the building). Therefore, CV might be able to optimize the wireless identification of LOS and NLOS, making it faster and more accurate with less radio resources for sounding and measurement reporting.

Wireless energy transfer is considered as an effective energy supply scheme for sensors and wearable devices in IoT[17]. However, in order to improve energy transfer efficiency, directional energy transfer based on beamforming requires more real-time and accurate channel feedback, which will reduce the energy utilization efficiency of RF powered devices[18]. The energy transmitter equipped with visual perception system can detect and locate potential energy receivers in line of sight, thus saving part of the channel sounding/feedback overhead.

As in cell edge area both control channels and traffic channels tend to suffer from poor coverage and/or strong inter-cell interference, visual positioning may be able to assist and enhance inter-cell interference coordination based on beamforming, further reducing inter-cell interference, and improving system capacity and throughput.

## VII. CONCLUSIONS

The exposure avoidance method is introduced in outdoor LOS scenario where large-scale antenna array are used in the millimeter wave band, of which the purpose is to reduce the exposure of human body in spatial dimension. Simulation results show that the method can reduce exposure. At the same time, more fine-grained beams are applied to balance exposure reduction and the loss of communication quality. The results show that finer beam granularity can achieve better link quality while reducing the EM exposure.

## ACKNOWLEDGMENT

This paper is sponsored by National Science and Technology Major Project of the Ministry of Science and Technology with grant No. 2018ZX03001024-006.